# Nonextensivity and the power-law distributions for the systems with self-gravitating long-range interactions

Du Jiulin[*]

*Department of Physics, School of Science, Tianjin University, Tianjin 300072, China*

**Abstract** By a natural nonextensive generalization of the conservation of energy in the $q$-kinetic theory, we study the nonextensivity and the power-law distributions for the many-body systems with the self-gravitating long-range interactions. It is shown that the power-law distributions describe the long-range nature of the interactions and the non-local correlations within the self-gravitating system with the inhomogeneous velocity dispersion. A relation is established between the nonextensive parameter $q \neq 1$ and the measurable quantities of the self-gravitating system: the velocity dispersion and the mass density. Correspondingly, the nonextensive parameter $q$ can be uniquely determined from the microscopic dynamical equation and thus the physical interpretation of $q$ different from unity can be clearly presented. We derive a nonlinear differential equation for the radial density dependence of the self-gravitating system with the inhomogeneous velocity dispersion, which can correctly describe the density distribution for the dark matter in the above physical situation. We also apply this $q$-kinetic approach to analyze the nonextensivity of self-gravitating collisionless systems and self-gravitating gaseous dynamical systems, giving the power-law distributions the clear physical meaning.

**PACS** number(s): 05.20.Dd; 98.10.+z; 51.10.+y; 05.90.+m

---

[*] Email Address: jiulindu@yahoo.com.cn

# 1. Introduction

As we know, almost all the systems treated in statistical mechanics with Boltzmann-Gibbs (BG) entropy have usually been extensive and this property holds for the systems with short-range interparticle forces. When we deal with the systems with long-range interpaticle forces such as Newtonian gravitational forces and Coulomb electric forces, where the nonextensivity holds, the BG statistics may need to be generalized for their statistical description (Cohen, 2002; Baranger, 2002; Plastino, 2004; Boon and Tsallis, 2005; Tsallis, *et al*, 2005; Tamarit and Anteneodo, 2005; Abe, *et al*, 2005; Plastino, 2005; Du 2004a, 2004b,2004c,2004c). In recent years, nonextensive statistics (Tsallis statistics or *q*-statistics) based on Tsallis' entropy has been proposed as a generalization of BG statistics. This new statistics is attracting a great of attention since it has been developed as a very useful tool for statistical description of the complex systems whose certain properties are beyond the scope of BG statistics due to, for instance, the long-range interactions (Gell-Mann and Tsallis, 2004; Abe and Okamoto, 2001; Boon and Tsallis, 2005; Tsallis, *et al*, 2005; Tamarit and Anteneodo, 2005; Abe, *et al*, 2005; Plastino, 2005). It has been applied extensively to deal with increasing varieties of interesting physical problems, among which self-gravitating systems and plasma systems offer the best framework for searching into the nonextensive effects because the long-range interactions between particles play a fundamental role in determining the properties of such systems (Lima, *et al*, 2002, 2000; Lavagno and Quarati, 2006; Du, 2004a, 2004b, 2004c, 2004d, 2005, 2006a; Leubner, 2004, 2005; Leubner and Voros, 2004).

A lot of works on the applications of the *q*-statistics to the fields of astrophysics have been made in recent years. However, in the light of present understanding, the statistical base suitable for the long-rang interacting systems, such as the self-gravitating system with inhomogeneous velocity dispersion, has still not been well established and, especially, the true physical nature of the nonextensive parameter *q* has not been well understood yet (Cohen, 2002; Baranger, 2002; Plastino, 2004; Boon and Tsallis, 2005; Tsallis, *et al*, 2005; Tamarit and Anteneodo, 2005; Abe, *et al*, 2005; Plastino, 2005). We need to know under what circumstances, *e.g.* which class of nonextensive systems and under what physical situation, should the *q*-statistics be used for their statistical description. It has been very important how to understand the physical meaning of *q* and how to determine this parameter from the microscopic dynamics of the systems with long-range interactions (for instance, the self-gravitating systems) in the *q*-statistics and its applications to astrophysics and other fields (Tsallis and Brigatti, 2004; Tsallis, 2004, Tsallis *et al*, 2002; Pluchino, *et al*, 2004; Almeida, 2001; Lavagno and Quarati, 2006, Du, 2004a, 2004b, 2004c, 2004d, 2005, 2006a; Shaikh, *et al*, 2006; Wuensche, *et al*, 2004; Wu and Chen, 2007).

Usually, the structure and stability of the self-gravitating many-body systems being at the statistical equilibrium are studied in terms of the maximization of the thermodynamic potential (BG entropy) under the constraints of fixed total mass and fixed total energy. This thermodynamic approach leads to Maxwell-Boltzmann (MB) distribution, an isothermal configuration that has been studied for a long time in the context of the self-gravitating systems such as stellar and galactic structure (Chandrasekhar, 1942; Binney and Tremaine, 1987). However, the isothermal sphere is found to be with infinite mass and infinite energy, thus exposed to contradiction with the requirement for finite mass and finite energy. Moreover, due to the long-range nature of the gravitational forces, the self-gravitating system is generally nonequlibrium and open. It often could reach to the hydrostatic equilibrium or the nonequilibrium stationary-state, but not to the state of thermal equilibrium. The isothermal configuration known about the self-gravitating systems is corresponding to the meta-stable locally mixing state only, not the true equilibrium state.

The kinetic theory of the systems with long-range inter-particle interactions has not been well understood yet. The main difficulties for the self-gravitating many-body systems are how rigorously to deal with the kinetics about the many-body collisions due to the long-range nature of the interparticle gravitational interactions. The simplest case of such examples is the two-body encounters (Binney and Tremaine, 1987): the motion of one particle is driven by the gravitational potential of the whole system and so the particle experiences quite a lot of two-body encounters. In a sense, each particle is constantly feeling the influences by all the other particles of the system. The long-range interactions and the non-local or global correlations within the system make the energy of particles not to be constant during the tow-body encounters. In other words, the total energy of the particles is nonextensive. While the total momentum of tow particles is also not conservation during the tow-body encounters because of presence of the external force field, which is different from the cases in the systems with short-range interparticle interactions. These facts inspire us to seek the new kinetic approach to describe the global correlations and the nonextensive thermodynamic properties of the self-gravitating many-body systems. The nonextensive $q$-statistics based on Tsallis' entropy may be relevant for the non-local description of such systems (Plastino 2005; Leubner 2005; Du, 2006a, 2006b, 2006c).

In this paper, we study the nonextensivity and the possible power-law distributions for the systems with self-gravitating long-range interactions from the $q$-kinetic theory based on Tsallis' entropy. Correspondingly, we determine the nonextensive parameter $q$ from the microscopic dynamical equations under the different astrophysical situations, so presenting the distribution functions the physical meanings. In sec 2, we deal with the generalized Boltzmann equation in the $q$-kinetic theory and discuss the necessary and sufficient condition for the solutions when the systems reach the $q$-equilibrium so as to derive the power-law distributions. In sec 3, we deal with the power-law distribution functions for three different astrophysical situations, where the nonextensivity of the self-gravitating many-body systems is studied by introducing the $q$-generalization of the conservation of energy. We derive the formula expressions of the nonextensive parameter $q$ and so present the power-law distributions clearly physical interpretations. Finally in sec 4, we give the summary and conclusions.

**2. The generalized Boltzmann equation and the $q$-kinetic theory**

To begin with, we briefly recall the standard kinetic theory in BG statistics. When we consider the system with $N$ particles interacting via the Newtonian gravitation $\mathbf{F}=-\nabla\varphi(\mathbf{r})$, if let $f(\mathbf{r},\mathbf{v},t)$ be the distribution function of the particles at time *t*, position **r** and with the velocity **v**, the dynamical behavior of the system can be described usually by the Boltzmann equation (Chapman and Cowling, 1970; Wang Zhuxi, 1978),

$$\frac{\partial f}{\partial t}+\mathbf{v}\cdot\frac{\partial f}{\partial \mathbf{r}}-\nabla\varphi\cdot\frac{\partial f}{\partial \mathbf{v}}=C(f), \tag{1}$$

where $\varphi$ is the potential and $C(f)$ is the collision term that, under the approximation of two-body collision, is given by

$$C(f)=\int\int(f'f_1'-ff_1)d\mathbf{v}_1\Lambda d\Omega, \tag{2}$$

By using the *H* theorem we get $C(f)=0$ and

$$\ln f+\ln f_1=\ln f'+\ln f_1' \tag{3}$$

when the system reaches the equilibrium state. The behavior of Eq.(3) is like a moving constant (conservation quantity). With the assumption that the total particle numbers, the total momentums and the total kinetic energy of the tow particles conserve during the tow-body collisions, we can derive the celebrated MB distribution function as the equilibrium distribution of the system, expressed by

$$f(\mathbf{r},\mathbf{v})=\left(\frac{m}{2\pi kT}\right)^{\frac{3}{2}}n(\mathbf{r})\exp\left(-\frac{m\mathbf{v}^2}{2kT}\right). \tag{4}$$

where *T* temperature is a constant and *m* is the mass of each particle. The density distribution is therefore given by $n(\mathbf{r})=n_0\exp[-m[\varphi(\mathbf{r})-\varphi_0]/kT]$, where $n_0$ and $\varphi_0$ are the density of particles and the gravitational potential at *r* =0, respectively. In the galactic dynamics, this equilibrium distribution is usually be written for the spherically symmetric stellar systems (Binney and Tremaine, 1987) as

$$f(\mathbf{r},\mathbf{v})=\frac{\rho_1}{(2\pi\sigma^2)^{3/2}}\exp\left(-\frac{\frac{1}{2}\mathbf{v}^2-\psi(\mathbf{r})}{\sigma^2}\right), \tag{5}$$

an isothermal configuration that has been studied for a long time in the context of the self-gravitating systems such as stellar and galactic structure, where $\sigma$ is called the velocity dispersion, formally corresponding to $\sigma = \sqrt{kT/m}$, $\psi(\mathbf{r}) = -\varphi(\mathbf{r}) + \varphi_0$ is called the relative potential, and $\rho_1$ is the density at $r = 0$. The gravitational potential should satisfy the Poisson's equation, *i.e.*

$$\nabla^2 \varphi = 4\pi G m n = 4\pi G \rho. \tag{6}$$

We now may introduce Tsallis' entropy and the nonextensive statistics. Tsallis' entropy (Tsallis, 1988) is

$$S_q = -k \sum_i p_i^q \ln_q p_i, \tag{7}$$

where $k$ is the Boltzmann constant, $p_i$ is the probability that the system under consideration is in its $i$th configuration such that $\sum_i p_i = 1$, $q$ is a real number whose deviation from unity is considered for measuring the degree of nonextensivity of the system, and the $q$-logarithmic function is defined by

$$\ln_q f = \frac{f^{1-q} - 1}{1 - q} \tag{8}$$

Correspondingly, the $q$-exponential function is defined by

$$e_q(f) = [1 + (1-q)f]^{1/1-q}, \tag{9}$$

BG entropy and the standard logarithmic function in BG statistics are recovered from them if we take $q \to 1$, $S_B = \lim_{q \to 1} S_q = -k \sum_i p_i \ln p_i$, and $\lim_{q \to 1} \ln_q f = \ln f$. The most distinguishing feature of the $q$-statistics is the nonextensivity (pseudoadditivity) for the entropy and the energy. If two compositions $A$ and $B$ are independent in the sense of factorization of the microstate probabilities with the energy spectrum $\{\varepsilon_i^A\}$ and $\{\varepsilon_j^B\}$, respectively, the Tsallis entropy and the energy spectrum corresponding to the composite system $A \oplus B$ can be written, respectively, by

$$S_q(A \oplus B) = S_q(A) + S_q(B) + (1-q)k^{-1} S_q(A) S_q(B) \text{ and}$$

$$\varepsilon_{ij}^{A \oplus B}(q) = \frac{1}{(1-q)\beta} \{1 - [1 - \beta(1-q)\varepsilon_i^A][1 - \beta(1-q)\varepsilon_j^B]\}, \tag{10}$$

where $\beta$ is the Lagrange parameters: $\beta = 1/kT$. Thus the entropy and the energy are both nonextensive in Tsallis $q$-statistics. When we take the $q \to 1$ limit, the extensivity (the standard additivity) of them in BG statistics is perfectly recovered.

In the $q$-kinetic theory based on the Tsallis entropy, if let $f_q(\mathbf{r},\mathbf{v},t)$ be the nonextensive distribution function of particles in the system with the long-range interacting potential $\varphi(\mathbf{r})$, Boltzmann equation, Eq.(1), can be generally generalized by the $q$-equation,

$$\frac{\partial f_q}{\partial t} + \mathbf{v} \cdot \frac{\partial f_q}{\partial \mathbf{r}} - \nabla \varphi \cdot \frac{\partial f_q}{\partial \mathbf{v}} = C_q(f_q), \tag{11}$$

where $C_q$ is called the $q$-collision term through which the nonextensivity effects can be incorporated. Under the assumption of tow-body collisions, it may be defined as

$$C_q(f_q) = \iint R_q(f_q, f_q') \Lambda d\mathbf{v}_1 d\Omega, \tag{12}$$

where (Lima et al, 2001)

$$R_q(f_q, f')_q = e_q(f'^{q-1} \ln_q f_q' + f_{q1}'^{q-1} \ln_q f_{q1}') - e_q(f^{q-1} \ln_q f_q + f_{q1}^{q-1} \ln_q f_{q1}).$$

In the limit $q \to 1$, we have $\lim_{q \to 1} R_q = f' f_1' - f f_1$, thus recovering the form in the standard kinetic theory. By using a nonextensive generalization of the "molecular chaos hypothesis" to Tsallis' statistics, it has been verified (Lima *et al*, 2001) that the solutions of the generalized Boltzmann equation (11) stratify the $q$-$H$ theorem and they converge irreversibly towards the equilibrium $q$-distribution, $f_q(\mathbf{r},\mathbf{v})$. The necessary and sufficient condition for such a $q$-equilibrium is

$$\ln_q f_q + \ln_q f_{q1} = \ln_q f_q' + \ln_q f_{q1}' \tag{13}$$

which determines the distribution function for the $q$-equilibrium. This implies that the sum of $q$-logarithms becomes the moving constant during the $q$-collisions. When $f_q(\mathbf{r},\mathbf{v})$ is determined by the condition Eq.(13), we have $C_q(f_q) = 0$ and Eq.(11) becomes

$$\mathbf{v}\cdot\frac{\partial f_q}{\partial \mathbf{r}} - \nabla\varphi\cdot\frac{\partial f_q}{\partial \mathbf{v}} = 0. \tag{14}$$

In other words, both the collisions and the movements would not lead to any variations of the *q*-distribution. This *q*-distribution function is stable. For the convenience to use, equivalently we can write Eq.(14) as

$$\mathbf{v}\cdot\nabla f_q^{1-q} - \nabla\varphi\cdot\nabla_v f_q^{1-q} = 0, \tag{15}$$

where we have denoted $\nabla = \partial/\partial\mathbf{r}$ and $\nabla_v = \partial/\partial\mathbf{v}$. Actually, Eq.(15) is similar to the Vlasov equation, which can also be derived directly by the Jeans' theorem (Binney and Tremaine, 1987). It should be kept in mind that any distribution functions that are considered for describing the equilibrium of the nonextensive system with long-range potential must satisfy Eq.(15).

## 3. The power-law distributions from the *q*-kinetic theory

We now apply Eq.(13) to the systems with self-gravitating long-range interactions to determine their distribution functions, then we use Eq.(15) to investigate the properties of the distribution functions and to determine the formula expression of the nonextensive parameter. With the *q*-kinetic theory, we firstly need to know who are the moving constants during the tow-body *q*-collisions. In the galactic dynamics, the collision is usually referred to the encounter since the true collision between particles is very few. The encounters are some slowly dynamical processes, representing the interparticle long-range interactions (While the usual collision is a rapid dynamical process, it plays a main role in the gaseous dynamics). For the self-gravitating systems, the total particle number conserves during the encounters, but the total momentum does not conserve because of presence of the external force field. Due to the self-gravitating long-range interactions and the non-local or global correlations within the system, generally speaking, the energy of particles behaves nonextensively. For example, the total energy of particles is not constant before and after the tow-body encounters, especially when the system is endowing a violent relaxation where the potential function varies with time (Binney and Tremaine, 1987).

## 3.1 The power-law distribution for the nonextensive system with self-gravitating long-range interactions

For the nonextensive system with the self-gravitating long-range interactions, in a sense, each particle is constantly feeling the influences by all the other particles in the system due to the long-range nature of gravitation. Additionally, if the system is endowing a violent relaxation, the potential function varies with time, which implies that only the particle number behaves as the moving constant, while the total energy and the total momentum are both not constant during the two-body encounters. We regard the two-body encounters as the tow-body $q$-collisions, the total energy of particles is nonextensive, and the nonextensive form of total energy can be considered as the moving constant during the tow-body $q$-collisions, i.e.

$$\varepsilon_q^{K+P} + \varepsilon_{q1}^{K+P} = \varepsilon_q'^{K+P} + \varepsilon_{q1}'^{K+P}, \quad (16)$$

Thus, from Eq.(13) we have

$$\ln_q f_q = a_0 + a_1 \varepsilon_q^{K+P}, \quad (17)$$

where $a_0$ and $a_1$ are the arbitrary coefficients, the upper index $K$ denotes the kinetic energy, $P$ the potential energy. The energy is taken as

$$\varepsilon_q^{K+P}(q) = \frac{1}{(1-q)\beta}\left\{1 - [1 - \beta(1-q)\frac{1}{2}mv^2][1 + \beta(1-q)m\psi]\right\}, \quad (18)$$

where we have used Eq.(10). The nonextensive form of total energy represents the long-range interparticle interactions and the non-local correlations within the system. For instance, two particles can be considered as two compositions before the encounter and they can be regarded as one "system" after the encounter, so the total energy of the "system" is nonextensive during the two-body encounter. While the nonextensive form of the energy can be expressed as Eq.(18) in terms of Eq.(10) because **v** and **r** are independent variables and then $K$ and $P$ are independent in the sense of factorization of the probabilities. Therefore, Eq.(16) is a naturally generalized form of the conservation of energy for the nonextensive system with the long-range interactions. In the limit $q \to 1$, we have $\varepsilon_1^{K+P} = \frac{1}{2}mv^2 - m\psi$, thus recovering the standard energy additivity.

Let $a_0 = (A_q^{1-q} - 1)/(1-q)$ and $a_1 = -\beta A_q^{1-q}$, from Eq.(17) we can derive a power-law distribution function,

$$f_q(\mathbf{r}, \mathbf{v}) = A_q \left[1 - (1-q)\beta \varepsilon_q^{K+P}\right]^{1/1-q}$$
$$= A_q \left[1 + (1-q)\beta m\psi\right]^{1/1-q} \left[1 - (1-q)\beta mv^2/2\right]^{1/1-q}. \tag{19}$$

Or, with the velocity dispersion $\sigma^2$ instead of $kT/m$, this distribution function can be written equivalently as

$$f_q(\mathbf{r}, \mathbf{v}) = \frac{B_q \rho_1}{(2\pi\sigma^2)^{3/2}} \left[1 + (1-q)\psi/\sigma^2\right]^{\frac{1}{1-q}} \left[1 - (1-q)v^2/2\sigma^2\right]^{\frac{1}{1-q}}, \tag{20}$$

where we have denoted $A_q = B_q \rho_1/(2\pi\sigma^2)^{3/2}$. $B_q$ is a $q$-dependent normalized constant. The velocity dispersion $\sigma$ is now a function of the space coordinate $\mathbf{r}$. From Eq.(20) we can get the density distribution function,

$$\rho = \rho_1 \left[1 + (1-q)\psi/\sigma^2\right]^{\frac{1}{1-q}} = \rho_1 [1 - (1-q)(\varphi - \varphi_0)/\sigma^2]^{\frac{1}{1-q}}. \tag{21}$$

In the limit $q \to 1$, the MB distribution Eq.(5) is perfectly recovered from Eq.(20). Thus, the MB distribution is generalized for $q \neq 1$ by the power-law one.

As we know, any functions considered as the equilibrium distribution of the self-gravitating system must satisfy Eq.(15). We now investigate the properties of the above distribution function and determine the corresponding nonextensive parameter $q$. For Eq.(20) we consider

$$\nabla f_q^{1-q} = \left[1 - (1-q)\frac{v^2}{2\sigma^2}\right] \left\{\left[1 - (1-q)\frac{\psi}{\sigma^2}\right]\nabla A_q^{1-q} - A_q^{1-q}(1-q)\nabla\left(\frac{\psi}{\sigma^2}\right)\right\}$$
$$+ A_q^{1-q}\left[1 - (1-q)\frac{\psi}{\sigma^2}\right](1-q)v^2\sigma^{-3}\nabla\sigma, \tag{22}$$

$$\nabla_v f_q^{1-q} = -A_q^{1-q}\left[1 - (1-q)\frac{\psi}{\sigma^2}\right](1-q)\sigma^{-2}\mathbf{v}. \tag{23}$$

Substituting them into Eq.(15) we have

$$\mathbf{v} \cdot \left[1-(1-q)\frac{v^2}{2\sigma^2}\right]\left\{\left[1-(1-q)\frac{\psi}{\sigma^2}\right]\nabla A_q^{1-q} - A_q^{1-q}(1-q)\nabla\left(\frac{\psi}{\sigma^2}\right)\right\}+$$

$$\mathbf{v} \cdot A_q^{1-q}\left[1-(1-q)\frac{\psi}{\sigma^2}\right](1-q)v^2\sigma^{-3}\nabla\sigma + \nabla\varphi \cdot A_q^{1-q}\left[1-(1-q)\frac{\psi}{\sigma^2}\right](1-q)\sigma^{-2}\mathbf{v} = 0. \quad (24)$$

In this equation, the sum of coefficients of the first power and the third power for the velocity **v** must vanish, respectively, because **v** and **r** are independent variables. Thus, we obtain the sum of coefficients of the first power of **v**,

$$\left[1-(1-q)\frac{\psi}{\sigma^2}\right]\nabla A_q^{1-q} - A_q^{1-q}(1-q)\nabla\left(\frac{\psi}{\sigma^2}\right) + (1-q)A_q^{1-q}\left[1-(1-q)\frac{\psi}{\sigma^2}\right]\frac{\nabla\varphi}{\sigma^2} = 0, \quad (25)$$

and the sum of coefficients of the third power of **v**,

$$-\frac{1-q}{2\sigma^2}\left\{\left[1-(1-q)\frac{\psi}{\sigma^2}\right]\nabla A_q^{1-q} - A_q^{1-q}(1-q)\nabla\left(\frac{\psi}{\sigma^2}\right)\right\}+$$

$$+(1-q)A_q^{1-q}\left[1-(1-q)\frac{\psi}{\sigma^2}\right]\sigma^{-3}\nabla\sigma = 0. \quad (26)$$

Combining Eq.(25) with Eq.(26), we find an exact formula expression between the nonextensive parameter 1-$q$, the velocity dispersion gradient and the gravitational acceleration:

$$(1-q)\nabla\varphi + 2\sigma\nabla\sigma = 0. \quad (27)$$

For spherically symmetric systems, it can be written as

$$1-q = -2\sigma\frac{d\sigma}{dr}\bigg/\frac{d\varphi}{dr} = -2\sigma\frac{d\sigma}{dr}\bigg/\frac{GM(r)}{r^2}. \quad (28)$$

These relations imply that $q$ is equal to unity if and only if $\nabla\sigma$ is equal to zero. So the power-law distribution function (20) with $q$ different from unity describes the nonequilibrium characteristic of the self-gravitating system. The above formula expressions relate the nonextensive parameter $q$ to the velocity dispersion and the gravitational acceleration, thus presenting $q \neq 1$ a clearly physical interpretation. It is clear that the values of (1-$q$) represent the characteristic of the long-range interactions and the non-local correlations within the system. If combining Eq.(27) with Poisson's equation, we can connect 1-$q$ with the mass density and the inhomogeneous velocity

dispersion by

$$\sigma \nabla^2 \sigma + (\nabla \sigma)^2 = -2\pi G(1-q)\rho. \tag{29}$$

So we find

$$1 - q = -\frac{\sigma \nabla^2 \sigma + (\nabla \sigma)^2}{2\pi G \rho}. \tag{30}$$

The values of *q* can be determined completely by the interesting measurable quantities: the velocity dispersion and the mass density; everything of its connections with the nonequilibrium characteristics of the system can be taken in at a glance. Therefore, Eq.(20) represents the properties of the non-local or the global correlations within the system with the self-gravitating long-range interactions when being at the nonequilibrium stationary-state.

Furthermore, we use Poisson's equation Eq.(6), $\nabla^2 \psi = -4\pi G \rho$, which can be expressed for spherically symmetric systems by

$$\frac{1}{r^2}\frac{d}{dr}\left(r^2 \frac{d\psi}{dr}\right) = -4\pi G \rho. \tag{31}$$

Combining Eq.(31) with the density distribution Eq.(21) and eliminating the potential function $\psi$, we can derive a second-order nonlinear differential equation for the radial density dependence of the self-gravitating system,

$$\frac{d^2\rho}{dr^2} + \frac{2}{r}\frac{d\rho}{dr} - \frac{q}{\rho}\left(\frac{d\rho}{dr}\right)^2 + \frac{2}{\sigma}(\frac{d\sigma}{dr})(\frac{d\rho}{dr}) + \frac{4\pi G}{\sigma^2}\rho^2\left(\frac{\rho}{\rho_1}\right)^{q-1} = 0. \tag{32}$$

This equation is exactly the same as that one proposed recently for the density distribution of dark matter if let $d\sigma/dr = 0$ (Leubner, 2005), thus generalizing Leubner's equation to the system with the inhomogeneous velocity dispersion. We may connect it with the density profile for a dark matter, which was found accurately to reproduce the density profiles generated by *N*-body and hydrodynamic simulations for the self-gravitating systems. If we let *q*=1, then Eq.(32) becomes

$$\frac{d^2\rho}{dr^2} + \frac{2}{r}\frac{d\rho}{dr} - \frac{1}{\rho}\left(\frac{d\rho}{dr}\right)^2 + \frac{4\pi G}{\sigma^2}\rho^2 = 0. \tag{33}$$

This is just the form in the case of the MB isothermal sphere ((Binney and Tremaine, 1987).

## 3.2 The power-law distribution for the self-gravitating collisionless system

If the particles are moving in a smooth mean potential $\psi(\mathbf{r})$ of the whole self-gravitating system, then they may be approximately assumed to be collisionless, in which the week interactions between the particles are actually ignored or neglected (Binney and Tremaine, 1987). For such a collisionless system, the $q$-collision term in Eq.(11) is believed to vanish. For this case, it is usually expected that the total energy of the particles is fixed and it behaves like the moving constant, *i.e.*

$$m(\tfrac{1}{2}v^2 + \psi) + m_1(\tfrac{1}{2}v_1^2 + \psi_1) = m(\tfrac{1}{2}v'^2 + \psi') + m_1(\tfrac{1}{2}v_1'^2 + \psi_1'). \tag{34}$$

Of course, the total particle number conserves, but the total momentum does not conserve because of presence of the external force field. Thus, from Eq.(13) we have

$$\ln_q f_q = a_0 + a_1 m(\frac{1}{2}v^2 - \psi), \tag{35}$$

where $a_0$ and $a_1$ are the arbitrary coefficients. Let $a_0 = (A_q^{1-q} - 1)/(1-q)$ and $a_1 = -\beta A_q^{1-q}$, we can derive the power-law distribution function,

$$f_q(\mathbf{r}, \mathbf{v}) = A_q \left[ 1 - (1-q)\frac{m}{kT}(\frac{v^2}{2} - \psi) \right]^{1/1-q}, \tag{36}$$

where $A_q$ is a $q$-dependent normalized parameter. As usual, if we use the velocity dispersion $\sigma^2$ instead of $kT/m$ in Eq.(36), it can be written equivalently by

$$f_q(\mathbf{r}, \mathbf{v}) = \frac{B_q \rho_1}{(2\pi\sigma^2)^{3/2}} \left[ 1 - (1-q)\frac{v^2/2 - \psi}{\sigma^2} \right]^{1/1-q}, \tag{37}$$

where we have denoted $A_q = B_q \rho_1 /(2\pi\sigma^2)^{3/2}$. It is clear that in the limit $q \to 1$ this distribution recovers the MB distribution correctly. The power-law distribution function, Eq.(37), is just that one presented for a collisionless stellar system by Lima and Souza (2005) in terms of a phenomenological analysis.

In Eq.(35), if let $a_0 = 1/(q-1)$ and $a_1 = A^{1-q}/(q-1)m$, we can write the power-law

distribution as

$$f_q(\mathbf{r}, \mathbf{v}) = A\left[-\frac{v^2}{2} + \psi\right]^{1/1-q}, \tag{38}$$

where $A$ is also a $q$-dependent normalized parameter. Therefore, Eq.(37) and Eq.(38) actually represent an identical $q$-equilibrium state. Now we may give them a clear microscopic dynamical meaning. Eq.(38) is just the distribution function derived in terms of maximizing Tsallis' entropy for a fixed total mass and fixed total energy ( Plastino and Plastino, 1993; Taruya and Sakagami, 2002), where $A$ was defined by $A = [(q-1)\beta/q]^{1/1-q}$. By a formal comparison between the distribution function (38) and the so-called stellar polytropic distribution, some authors often connected the nonextensive parameter $q$ with the polytrope index $n$ using the relation $n = \frac{3}{2} + 1/(q-1)$. And just for this reason, many authors have often called this distribution the stellar polytropes (Chavanis and Sire, 2005; Sakagami and Taruya, 2004; Chavanis, 2006). However, we will prove here that such an understanding may be incorrect, and Eq.(38) or (37) is not the polytropic distribution but the *Tsallis isothermal spheres*.

The properties of the distribution functions, (37) or (38) can be studied by using Eq.(15) because any distribution functions used for describing the equilibrium of the self-gravitating system must satisfy this equation. For Eq.(37), we may consider

$$\nabla f_q^{1-q} = -\left[\frac{B_q \rho_1}{(2\pi)^{3/2}}\right](1-q)\sigma^{-3(1-q)-1}\left[3\nabla\sigma - \frac{\nabla\psi}{\sigma} - (5-3q)\frac{\nabla\sigma}{\sigma^2}(\frac{v^2}{2} - \psi)\right], \tag{39}$$

$$\nabla_v f_q^{1-q} = -\left[\frac{B_q \rho_1}{(2\pi)^{3/2}}\right](1-q)\sigma^{-3(1-q)-2}\mathbf{v}. \tag{40}$$

Substitute them into Eq.(15) we have

$$-\mathbf{v}\cdot\left[3\nabla\sigma - \sigma^{-1}\nabla\psi - (5-3q)\sigma^{-2}\nabla\sigma(\frac{v^2}{2} - \psi)\right] + \sigma^{-1}\mathbf{v}\cdot\nabla\varphi = 0. \tag{41}$$

Let the sum of coefficients for the first power and the third power of the velocity $\mathbf{v}$ vanish, respectively, we find, for any values of $q$,

$$\nabla\sigma = 0. \tag{42}$$

This implies that the distribution function (37) represents the isothermal processes in the system. Also, if we substitute the distribution function (38) into Eq.(15), we have, for any values of $q$,

$$\nabla A = 0, \text{ i.e. } \nabla T = 0. \tag{43}$$

Thus Eq.(38) also represents the isothermal processes in the system.

The above analyses show that the power-law distribution functions, (37) and (38), are both *the Tsallis isothermal spheres* for any $q \neq 1$, and when $q=1$ they become the MB isothermal sphere. So, strictly speaking, the power-law distribution function (38) is not in agreement with the polytropic state we have usually known, because the polytropic process is a *nonisothermal* ones with the various polytropic index $n$, only when the index $n \rightarrow \infty$ is it isothermal. Thus from the $q$-kinetic theory we can derive the power-law distribution functions, (37) and (38), of the self-gravitating collisionless systems, but also obtain the clearly understanding of their physical properties.

### 3.3 The power-law distribution for the self-gravitating gaseous dynamical system

For the self-gravitating gaseous dynamical systems, such as the matter inside a normal star, instead of the encounter, the collisions between the particles play a main role in the gaseous dynamics. The collisions are the rapid dynamical processes. While the nonextensivity effects introduced by the encounters (*i.e.* slow dynamical processes in the gas with interparticle long-range interactions) can be incorporated as the $q$-collision in the nonextensive $q$-kinetic theory. For this case, the total particle number, the total moment and the total kinetic energy conserve during the tow-body $q$-collisions and they behave like the moving constants (Chapman and Cowling, 1970; Wang Zhuxi, 1978), *i.e.*

$$m\mathbf{v} + m_1\mathbf{v}_1 = m\mathbf{v}' + m_1\mathbf{v}'_1,$$

$$\tfrac{1}{2}mv^2 + \tfrac{1}{2}m_1v_1^2 = \tfrac{1}{2}mv'^2 + \tfrac{1}{2}m_1v_1'^2. \tag{44}$$

So, from Eq.(13) we have

$$\ln_q f_q = a_0 + \mathbf{a}_1 \cdot m\mathbf{v} + a_2 \frac{1}{2}mv^2, \tag{45}$$

where $a_0$ and $a_2$ are arbitrary coefficients and $\mathbf{a}_1$ is an arbitrary constant vector. If let

$a_0 = (B^{1-q} - 1)/(1-q) - \frac{1}{2} m\beta v_0^2 B^{1-q}$, $\mathbf{a}_1 = -\beta B^{1-q} \mathbf{v}_0$ and $a_2 = -\beta B^{1-q}$, then we can derive the power-law distribution function in the familiar form,

$$f_q(\mathbf{r}, \mathbf{v}) = B\left[1 - (1-q)\frac{m(\mathbf{v} - \mathbf{v}_0)^2}{2kT}\right]^{1/1-q}, \tag{46}$$

where $\mathbf{v}_0$ is the barycentric velocity of the system, $B = B_q \rho (m/2\pi kT)^{3/2}$ and $B_q$ is a $q$-dependent normalized constant. Formally the distribution function (46) is the same as the $q$-Maxwell's distribution proposed by Silva *et al* (1998), but now the density $\rho$ and temperature $T$ are dependent on the space coordinate $\mathbf{r}$. So, Eq.(46) generalizes the MB distribution to the inhomogeneous case in the $q$-kinetic theory.

To study the properties of the power-law distribution (46) and determine the corresponding nonextensive parameter $q$, we may consider

$$\nabla f_q^{1-q} = \left[1 - (1-q)\frac{m(\mathbf{v} - \mathbf{v}_0)^2}{2kT}\right]\nabla B^{1-q} - B^{1-q}(1-q)\frac{m(\mathbf{v} - \mathbf{v}_0)^2}{2k}\nabla\left(\frac{1}{T}\right),$$

$$\nabla_v f_q^{1-q} = -B^{1-q}(1-q)\frac{m(\mathbf{v} - \mathbf{v}_0)}{kT}, \tag{47}$$

and substitute them into Eq.(15),

$$\left[1 - (1-q)\frac{m(\mathbf{v} - \mathbf{v}_0)^2}{2kT}\right]\mathbf{v} \cdot \nabla B^{1-q} - B^{1-q}(1-q)\frac{m(\mathbf{v} - \mathbf{v}_0)^2}{2k}\mathbf{v} \cdot \nabla\left(\frac{1}{T}\right)$$

$$+ B^{1-q}(1-q)\frac{m(\mathbf{v} - \mathbf{v}_0)}{kT} \cdot \nabla\varphi = 0. \tag{48}$$

In this equation, the sum of coefficients for the zeroth, the first, the second and the third power of the velocity $\mathbf{v}$ must vanish, respectively. For the coefficients of the first power of $\mathbf{v}$, we have

$$\nabla B^{1-q} - (1-q)\frac{m\mathbf{v}_0^2}{2kT}\left[\nabla B^{1-q} + B^{1-q}T\nabla\left(\frac{1}{T}\right)\right] + B^{1-q}(1-q)\frac{m\nabla\varphi}{kT} = 0, \tag{49}$$

For the coefficients of the second and the third power of $\mathbf{v}$, we have

$$\nabla B^{1-q} + B^{1-q}T\nabla\left(\frac{1}{T}\right) = 0, \tag{50}$$

For the terms of the zeroth power of **v**, we find

$$\mathbf{v}_0 \cdot \nabla \varphi = 0. \tag{51}$$

Substituting Eq.(50) into Eq.(49) we get

$$\nabla B^{1-q} + B^{1-q}(1-q)\frac{m\nabla \varphi}{kT} = 0. \tag{52}$$

Combining Eq.(52) with Eq.(50), we exactly find a relation between the nonextensive parameter $q$, the temperature gradient and the gravitational acceleration:

$$k\nabla T + (1-q)m\nabla \varphi = 0. \tag{53}$$

The fact that $q$ is different from unity if and only if $\nabla T \neq 0$ is got at a glance. This formula expression for $q$ was derived firstly in 2004 (Du, 2004a, 2004b). And then it led to a test of nonextensive statistics by the solar sound speed measured in helioseismology (Du, 2006b). From the above $q$-kinetic analyses we clearly know which class of nonextensive systems and under what physical situations the generalized MB distribution (46) is suitable for their statistical description.

Applying Poinsson's equation to Eq.(53) we easily get

$$1 - q = -k\nabla^2 T / 4\pi G\rho m. \tag{54}$$

Eq.(53) and Eq.(54) relate closely $q \neq 1$ to the non-local or global characteristics of the gaseous dynamical system with the self-gravitating long-range interactions when it is in the nonequilibrium stationary-state, thus presenting a clearly physical interpretation for the parameter $q$ different from unity.

The other property of the power-law distribution (46) can be analyzed from the condition Eq.(51). This condition gives a constraint to $\mathbf{v}_0$, the macroscopic entirety-moving velocity of the gaseous system, when the system reaches the nonequilibrium stationary-state. It requires that $\mathbf{v}_0$ should always be vertical to the gravitational force.

Furthermore, from Eq.(52) we can derive the density distribution,

$$\rho(\mathbf{r}) = \rho_1 \left(\frac{T}{T_0}\right)^{3/2} \exp\left[-\frac{m}{k}\int \frac{\nabla \varphi}{T} \cdot d\mathbf{r} + \frac{m\varphi_0}{kT_0}\right], \tag{55}$$

where $\rho_1$, $T_0$, $\varphi_0$ are the integral constants, they denote the density, the temperature and the potential at $r=0$, respectively. If $q=1$ we have $\nabla T = 0$ and $T = T_0 =$ constant, and then Eq.(55) recovers the density profiles in the standard MB distribution. Thus, Eq.(46) generalizes the MB distribution in the $q$-kinetic theory for the nonequilibrium self-gravitating gaseous system with the non-local characteristics: $\nabla T \neq 0$.

## 4. Summary and conclusions

In this article, we apply the nonextensive $q$-kinetic theory based on Tsallis' entropy to study the power-law distributions for the many-body systems with self-gravitating long-range interactions. We first deal with the generalized Boltzmann equation in the $q$-kinetic theory that may be used for governing the self-gravitating system. On the basis of the $q$-H theorem, the solutions of the generalized Boltzmann equation can converge irreversibly towards the $q$-equilibrium distribution function, $f_q(\mathbf{r}, \mathbf{v})$, therefore they must satisfy Eq.(13), a necessary and sufficient condition for this $q$-equilibrium of the system. Eq.(13) implies that the sum of $q$-logarithms is the moving constant during the two-body $q$-collisions. In the light of the different physical situations of the system under consideration, we are able to derive the $q$-equilibrium distribution functions, the $q$-dependent power-law distributions. Furthermore, by using the microscopic dynamical equation, Eq.(15), the properties of the power-law distribution functions can be precisely analyzed, the formula expressions of the corresponding nonextensive parameter $q$ can been uniquely determined, and so the physical interpretations of $q$ different from unity can be clearly presented.

Under three different physical situations for the self-gravitating systems, we deal with the nonextensive power-law distributions in the $q$-kinetic theory.

Firstly, we consider the self-gravitating many-body system where the two-body encounters play a main role in the microscopic dynamics. Due to the long-range nature of the self-gravitating interactions and the non-local or global correlations within the system, in a sense, each particle is constantly feeling the influences by all the other particles in the system. The energy of particles behaves nonextensively. Thus, only the particle number is the moving constant during the two-body encounters, while the total energy and the total momentum are both not the moving constant. For such a case, the total energy of particles is nonextensive. The encounters can be regarded as the *q*-collisions and the nonextensive form of the energy can be considered as the moving constant during the two-body *q*-collisions. So we obtain the power-law distribution function, Eq.(20), for the nonextensive system with self-gravitating long-range interactions. Furthermore, using the microscopic dynamical equation, Eq.(15), we study the physical properties of the power-law distribution function and obtain the formula expression of the *q* parameter, Eq.(27). Moreover, by the formula (30) we establishes a relation between the nonextensive parameter *q* and the interesting measurable quantities: the velocity dispersion gradient and the mass density of the system, thus presenting $q \neq 1$ a clearly physical interpretation. We show that Eq.(20) generalizes the MB distribution to the physical situation of the system with the inhomogeneous velocity dispersion (*i.e.* $d\sigma/dr \neq 0$) and so correctly describes the non-local or global correlations within the self-gravitating many-body system being in the nonequilibrium stationary-state.

Using Poisson's equation, we also derive a second-order nonlinear differential equation for the radial density dependence of the self-gravitating system, Eq.(32), which can describe the radial density distribution of the dark matter under the physical situation of $d\sigma/dr \neq 0$. It was found to be able to reproduce accurately the density profiles generated by *N*-body and hydrodynamic simulations for the self-gravitating systems. When *q*=1, this equation correctly reduces to the case of MB isothermal sphere.

Secondly, we study the so-called self-gravitating collisionless system, where the total energy of the particles is fixed during their moving in the mean potential $\psi(\mathbf{r})$ of the whole system. For this case, the particle number and the total energy behave like the moving constant. We derive the power-law distribution function Eq.(37) or, equivalently, Eq.(38). It is shown that this power-law distribution function describes the physical situation of *Tsallis isothermal sphere*, which is with the homogeneous velocity dispersion: $\nabla \sigma = 0$ for the arbitrary values of $q \neq 1$. It is not the stellar polytropic as called by some authors because Eq.(38) is not in agreement with the properties of polytropic state. When we take $q=1$, the MB isothermal sphere is recovered from the power-law distribution.

Thirdly, We revisit the self-gravitating gaseous dynamical system, where the interparticle collisions play a main role in the gas dynamics. While the nonextensivity effects introduced by the interparticle long-range interactions can be incorporated as the $q$-collision. For such a case, we think that the total particle number, the total moment and the total kinetic energy conserve during the tow-body $q$-collisions and they behave like the moving constant, so the power-law distribution function, Eq.(46), is obtained, which is shown to represent the non-local or global correlations within the self-gravitating gas being in the nonequlibrium stationary-state with $\nabla T \neq 0$.

All the power-law distribution functions for the systems with self-gravitating long-range interactions can be derived by the condition (13) if only we have known or determined the moving constants in accordance with the different physical situations of the interacting particles in the systems. Furthermore, the formula expressions of the nonextensive parameter $q$ different from unity for the various cases can be determined by using the dynamical equation (15), and therefore their physical meanings can be clearly presented.

I would like to thank the National Natural Science Foundation of China under the grant No.10675088 and the "985" program of TJU for the financial supports.